\documentclass[12pt]{article}

\usepackage{latexsym}
\usepackage{graphicx}\usepackage{epsfig}
\newcommand{\beq}{\begin{eqnarray}}
\newcommand{\eeq}{\end{eqnarray}}

\def\e{{\rm e}}
\def\t{\widetilde}

\newcommand{\eg}{{\it e.g.,}\ }
\newcommand{\ie}{{\it i.e.,}\ }
\newcommand{\reef}[1]{(\ref{#1})}

\begin {document}

\thispagestyle{empty}

\begin{titlepage}
\begin{flushright}
{\small
}
\end{flushright}
\vskip 1cm

\centerline{\Large \bf Fundamental Plasmid Strings}
\vskip 0.3cm
\centerline{\Large \bf and}
\vskip 0.3cm
\centerline{\Large \bf Black Rings}

\vspace{0.5cm}

\vskip 0.7cm
\centerline{ 
Jose J. Blanco-Pillado$^{a,}$\footnote{E-mail:
jose@cosmos.phy.tufts.edu},
Roberto Emparan$^{b,}$\footnote{E-mail: emparan@ub.edu}
and Alberto Iglesias$^{c,}$\footnote{E-mail: iglesias@physics.ucdavis.edu}}

\vskip .5cm
\centerline{$^a$\em Institute of Cosmology,
Department of Physics and Astronomy,}
\centerline{\em Tufts University, Medford, MA 02155  }
\bigskip
\centerline{$^b$\em Instituci\'o Catalana de Recerca i Estudis
Avan\c cats (ICREA),}
\centerline{\em Passeig Llu{\'\i}s Companys, 23, 08010 Barcelona, Spain} 
\centerline{\em and}
\centerline{\em Departament de F{\'\i}sica Fonamental, Universitat de
Barcelona,}
\centerline{\em Diagonal 647, E-08028 Barcelona, Spain}
\bigskip
\centerline{$^c$\em Department of Physics, University of California, Davis,
CA 95616}

\vskip 1cm

\vskip 1cm

\begin{abstract}
 
\noindent We construct excited states of fundamental strings that admit
a semiclassical description as rotating circular loops of string. We
identify them with the supergravity solutions for rotating dipole rings.
The identification involves a precise match of the mass, radius and
angular momentum of the two systems. Moreover, the degeneracy of the
string state reproduces the parametric dependence of the entropy in the
supergravity description. When the solutions possess two macroscopic
angular momenta, they are better described as toroidal configurations
(tubular loops) instead of loops of string. We argue that the decay of
the string state can be interpreted as superradiant emission of quanta
from the ergoregion of the rotating ring. 

\end{abstract}
 
\end{titlepage}

\setcounter{footnote}{0}

\section{Introduction}

Fundamental strings contain excited states that admit a semiclassical
description. These can be regarded as macroscopic strings, and
are of interest for many reasons. On the one hand, since they are built
using perturbative string theory, many of their properties can be
understood in detail. On the other hand, being macroscopic objects, they
can source spacetime fields and give rise to supergravity solutions of
string theory at low energy. Such dual descriptions, first advanced in
\cite{Dabholkar:1989jt,Dabholkar:1990yf,Dabholkar:1995nc,Callan:1995hn,
Sen:1995in}, are at the basis of many recent developments in the
microphysics of black holes, \eg \cite{Dabholkar:2004yr}, and novel
AdS/CFT dualities \cite{Dabholkar:2007gp,Lapan:2007jx,Kraus:2007vu}. 
Semiclassical
string states also provide a handle on certain solvable regimes of the
gauge/string correspondence \cite{Gubser:2002tv}. Additionally, they may
play a role in cosmology, see \eg \cite{Copeland:2003bj}.

In this paper we are mainly concerned with the use of semiclassical
string states for the microscopic interpretation of black hole-like
objects in supergravity. Specifically, we make a connection between a
class of states of fundamental strings, namely circular loops of string,
following \cite{BlancoPillado:2005px,BlancoPillado:2007hi}, and the
supergravity solutions that describe black rings with a dipole of the
Kalb-Ramond field \cite{Emparan:2004wy,Emparan:2006mm}. 
These loops of fundamental string do not possess any conserved
gauge charges, and they are not supersymmetric. They also possess
angular momentum: the loop is rotating, and its centrifugal repulsion
prevents its collapse. 

\begin{figure}
\centerline{\includegraphics[width=10cm]{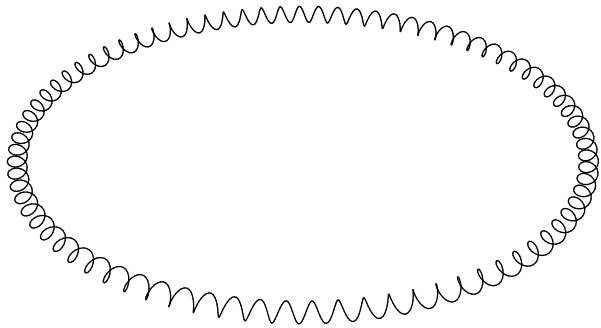}}
\caption{\small Plasmid string, describing a rotating loop
of string. The rotation of the ring corresponds to the helical advance
of the string. For a generic rotating string loop, the profile of the
string oscillations is not a circular helix but a wiggly structure.}
\label{fig:plasmid}
\end{figure}

The rigid rotation of a fundamental string is not possible because of
reparametrization invariance of the worldsheet, but a loop of excited
fundamental string can rotate. An example involving fermionic
excitations was presented in \cite{BlancoPillado:2005px}; here we will
discuss a simpler construction using bosons. The essence of the idea is
captured pictorially by the notion of a {\em plasmid} string
(fig.~\ref{fig:plasmid})\footnote{A plasmid is a circular molecule of a
double-stranded DNA helix. The fundamental plasmid string is instead
single-stranded.}: a helical string that closes in on itself on a
circle, with the helical advance of the string resulting in the coherent
rotation of the state along the ring circle. For a generic state the
oscillations of the string do not have the profile of a circular helix,
but are replaced by small-scale wiggles of the string whose propagation
along the circular loop give the ring its angular momentum. They also
give rise to a large degeneracy of the macroscopic loop. 

Our identification of the microscopic string state and the supergravity
solution is accurate when the ring is {\em thin}, in the sense that its
radius is large and its self-gravitation is weak. In this regime we are
able to precisely match the relations between the mass, angular
momentum, radius and dipole charge (\ie the winding $n_w$ around the
ring circle) that appear in both sides of the equivalence.

Properly speaking, the solutions we study are not black, since they have
naked singularities instead of regular horizons. However, in the spirit
of \cite{Sen:1995in}, one can associate a stretched horizon to these
solutions, and as we shall see, the area of this horizon reproduces
the degeneracy of the string state up to an undetermined numerical
coefficient. In fact a regular horizon is expected to arise when
higher-derivative corrections to the action are included---they are then
{\em small} black rings. The possible connection between these string
states and black rings was anticipated in \cite{Emparan:2004wy}.

When the helix radius becomes macroscopically large, which happens when
the angular momentum on the plane orthogonal to the plane of the ring is
macroscopic, the configuration is not adequately described as a ring,
but rather as a tubular loop. In fact we shall argue that it is a loop
of (a certain) supertube. These configurations can be analyzed in the
approximation in which the radius of the loop is very large, and they
reproduce the properties of the corresponding string states.

None of these string states is supersymmetric and therefore they are
expected to decay once string interactions are turned on. We shall argue
for an interpretation of the decay in terms of supergravity: it
corresponds to the spontaneous emission of superradiant modes from the
ergoregion that surrounds the ring. The quantitative development of this
idea is technically somewhat involved and is left for the future, but at
least we are able to provide a qualitative description of the
equivalence using the ideas in \cite{Emparan:2007en,DEM}.

The paper is divided into three main sections, each one describing a
different construction of the loop of string:
section~\ref{sec:classical} builds it as a classical solution to the
Nambu-Goto equations; section~\ref{sec:micro} obtains it as a
quantum state of string theory; section~\ref{sec:sugra} then compares
these configurations with the supergravity solution for a dipole ring.
The last section comments on some consequences and possible extensions
of these results.

\setcounter{equation}{0}
\section{Classical Fundamental Plasmid String}
\label{sec:classical}

In this section we present solutions of the classical equations of
motion for a fundamental bosonic string describing circular loops
stabilized by angular momentum, \ie plasmid strings. The lowest
dimension in which our construction can be made is $D=5$, and for
clarity we will only work out explicitly this case. The
extension to higher dimensions is nevertheless straightforward.

Our starting point is the Nambu-Goto action for a string in
five-dimensional flat space,
\beq
S_{NG} = -\frac{1}{2 \pi \alpha'} \int{\sqrt{-\gamma}\,d\sigma\,d\tau}\,,
\eeq
where $1/(2\pi\alpha')$ is the tension of the string, $\gamma$ denotes 
the determinant of the induced metric on the worldsheet, 
$\gamma_{m n} = \eta_{\mu \nu} \partial_m X^{\mu}\partial_n
X^{\nu}$, and $m,n = 0,1$, $\mu,\nu = 0,\dots,4$ are the worldsheet and
spacetime indices respectively. The equations of motion obtained from
this action are
\beq
\partial_m (\sqrt{-\gamma} \gamma^{mn} \partial_n X^{\nu})= 0\,.
\eeq
It is convenient to rewrite these equations in the
conformal gauge where $\gamma_{mn} = \sqrt{-\gamma}\,\eta_{mn}$. In
this gauge the equations of motion are just the two-dimensional wave
equation,
\beq
\partial^2_{\sigma} X^{\mu} - \partial^2_{\tau} X^{\mu}= 0\,.
\eeq
Furthermore, we can use the residual gauge freedom to fix 
$X^0= p_0\tau$, so the previous equations become
\beq
\partial^2_{\sigma} X^{i} - \partial^2_{\tau} X^{i}= 0\,,
\eeq
where $X^i$ denotes the four-dimensional spatial vector describing
the position of string in our five-dimensional spacetime. The most
general solution of these equations is
\beq
X^i = \frac {1} {2} [A^i(\tau - \sigma) + B^i(\tau + \sigma)]\,.
\eeq
Finally we still have to make sure that our solution satisfies the 
constraints, which in our gauge impose the conditions
\beq
\label{constraints}
|\partial_{\sigma} A^i|^2 = |\partial_{\sigma} B^i|^2 = p_0^2.
\eeq

Armed with the most general solution, we can now try to look for
configurations of the type we are interested in, namely, stationary circular
loops stabilized by angular momentum. At first, this sounds impossible
to do. A fundamental bosonic string does not have any longitudinal
degrees of freedom since any such apparent motion can always be
compensated by a change of gauge. On the other hand, we can imagine the
situation in which small-scale wiggles propagate along the
string in a nearly circular loop, making it possible for it to 
have some angular momentum perpendicular to the ring. The travelling 
wiggles produce a centrifugal force that balances the tension of the 
string and therefore allow for stationary configurations. A family of 
such solutions can be easily written in the conformal gauge as
\begin{eqnarray}\label{plasmid}
X^0&=& p_0\tau\,,\nonumber\\
X^1&=& R  \cos \left(2n_w \left(\tau - \sigma\right)\right)\,,\nonumber\\
X^2&=& R  \sin \left(2n_w \left(\tau - \sigma\right)\right)\,,\nonumber\\
X^3&=& \frac{R\,n_w}{N} \cos \left(2 N\left(\tau + \sigma\right)\right)\,,\\
X^4&=& \frac{R\,n_w}{N} \sin \left(2 N\left(\tau + \sigma\right)\right)\nonumber\, ,
\end{eqnarray}
with $p_0=4n_wR$.
Taking $0\leq\sigma<\pi$, the solution represents a circular loop 
of radius $R$, winding $n_w$ times on the $X^1$-$X^2$ plane, and stabilized
against collapse by a chiral excitation of $N$ helical turns of 
amplitude $R n_w/N$ on the $X^3$-$X^4$ plane. 
The winding number $n_w$ is not topologically conserved since the string,
which is dynamically stabilized by rotation and not by topology, 
can be continuously shrunk to zero radius. 

It is straightforward to see 
that these solutions indeed fulfill the constraints described 
above, since both $\partial_{\sigma} A^i$ and $\partial_{\sigma} B^i$ 
parametrize circles of radius $p_0$. We have chosen these circles 
to lay on perpendicular planes in order to avoid the presence of 
cusps in these solutions. There are, of course, many other 
solutions similar to this one. In particular, we see that the
macroscopic shape of the loop can be arbitrary since the only constraint
on the function $A^i(\tau - \sigma)$ is given by Eq.(\ref{constraints}), which 
basically fixes the parametrization of $A^i$ but not its shape. 
The reason for this is that the balance between the tension and 
the centrifugal force produced by the wiggles is a local phenomenon 
that ensures that at each point on the string the effective tension 
vanishes.\footnote{Solutions of this type 
are known to exist in chiral superconducting string models 
\cite{BlancoPillado:2000ep} where the role of the chiral wiggles 
is played by a neutral current on the string worldsheet and the
strings states are the so-called chiral vortons \cite{Davis:1988jq}.} Here, 
however, we only focus on the circular string case. These solutions have been 
previously considered, for a different
purpose, in \cite{Frolov:2003qc,Arutyunov:2003uj}.

As we anticipated, the solutions presented above have a non-vanishing 1-2 component
of the angular momentum, which depends on the winding number as
well as the radius of the loop as
\begin{eqnarray}\label{J12nR}
J_{1 2} &=& \frac{1}{2 \pi \alpha'}\int_0^{\pi}{ d\sigma\, \left( X^1 \partial_{\tau} X^2
  -  X^2  \partial_{\tau} X^1\right)} = \frac{n_w R^2}{\alpha'}\,.
\eeq

On the other hand, we can see that $N$ not only controls the amplitude 
of the small oscillations but also enters in the calculation of the
3-4 component of the angular momentum,
\begin{eqnarray}
J_{3 4} &=& \frac{1}{2 \pi \alpha'}\int_0^{\pi}{ d\sigma\, \left( X^3 \partial_{\tau} X^4
  -  X^4  \partial_{\tau} X^3\right)} = \frac{n_w^2 R^2}{\alpha' N}\,.
\eeq

For a very large value of $N$ we find ourselves with an almost perfectly
circular string loop with only $J_{1 2}$ angular momentum. This is the
kind of solution we want to identify with our black ring solutions in
the supergravity description\footnote{The case $n_w=1$, $N=1$ was
presented in \cite{Chialva:2003hg}, but it will become clear that we do
not expect it to correspond to a black ring.}.

An interesting point about this gauge is that $\sigma$
parametrizes the energy along the string, so we can
easily read off the total of energy, \ie the mass, of a particular configuration 
from the total range of $\sigma$. In our example the mass
is
\beq\label{MRn}
M =\frac{1}{2 \pi \alpha'}\int_0^{\pi}p_0 d\sigma
 = \frac{2 R n_w}{\alpha'}\,.
\eeq
{}From the results above we also obtain
\beq\label{J12M}
J_{1 2} = M^2 \frac{\alpha'}{4 n_w}.
\eeq
This is always below the Regge bound $J=M^2\alpha'$, even when $n_w=1$ .
Hence we expect the existence of many other string configurations that
satisfy \reef{J12M} for given $M$, $J_{12}$ and $n_w$. This is, there must
exist a large degeneracy of these states. It appears that the origin of this
degeneracy is the possibility of varying the small-scale
structure of the loop without altering the relations \reef{MRn} and \reef{J12M}.

This is easily seen to be the case. The plasmid solution above is only one of the
simplest possible configurations with these properties but we can write
the most general solution for a circular ring with chiral perpendicular
travelling wiggles as
\begin{eqnarray}
\label{general}
X^0&=&4 n_wR\tau\,,\nonumber\\
X^1&=& \frac {1} {2} A^1 (\tau  - \sigma) = R \cos \left(2n_w\left(\tau - \sigma\right)\right)\,,\nonumber\\
X^2&=& \frac {1} {2}  A^2(\tau -\sigma) = R  \sin \left(2n_w\left(\tau - \sigma\right)\right)\,,\\
X^3&=& \frac {1} {2}  B^3 (\tau + \sigma)\nonumber\,,\\
X^4&=&  \frac {1} {2} B^4 (\tau + \sigma)\nonumber\,.
\end{eqnarray}
The general form for the two-dimensional vector $B^i (\tau +
\sigma)=B^i(\sigma_+)$ consistent with a closed string loop is given by
\beq
B^i(\sigma_+) =2 R n_w\,\sum_{n=1}^{\infty} {\left[ c^i_n 
\cos\left(2 n \sigma_+\right) + d^i_n \sin\left(2 n
  \sigma_+\right) \right]}\,,
\eeq
or, in terms of complex coefficients $\beta^i_n$, by
\beq
B^i(\sigma_+) =  2 i R n_w\,\sum_{n\not = 0}  {1\over n} \beta_n^i{\rm e}
^{-2in\sigma_+}\,,
\eeq
where 
\beq
\beta^i_n = -\frac{in}{2}(c^i_n + i d^i_n)\,,
\eeq
and $\beta_{-n} = \beta_n^*$. The only other constraint on these
functions comes from Eq.~(\ref{constraints}) which, in turn, imposes the following 
conditions on the expansion coefficients
\beq
\sum_{n\not = 0} \beta_n^i \beta^i_{m-n} = \delta_{m0}\,.
\label{Virasoro}
\eeq
Therefore any solution of the form (\ref{general}) that fulfills the
requirements of eq.~(\ref{Virasoro}) represents a nearly circular loop
with the same radius, mass and $J_{12}$ as the simple plasmid string
\reef{plasmid}. On the other hand, these
solutions have in general different values of $J_{34}$, 
\beq
J_{34} = \frac {R^2 n_w^2}{\alpha'}\sum_{n=1}^{\infty}{n\,(c^3_n d^4_n
  - d^3_n c^4_n)} = 
\frac {R^2 n_w^2}{\alpha'}\sum_{n=1}^{\infty}{n\,({\bf c}_n \times {\bf d}_n})\,.
\eeq
We can now obtain, following an argument parallel to one presented
in \cite{Lunin:2001fv}, an upper bound for this component of the
angular momentum. Observe that
\begin{eqnarray}
J_{34}^2 &=& \left(\frac {R^2 n_w^2}{\alpha'}\right)^2 
\sum_{m,n}{n\,m\,({\bf c}_n  \times {\bf d}_n) \cdot ({\bf c}_m \times
  {\bf d}_m)} \nonumber\\
&=& \left(\frac {R^2 n_w^2}{\alpha'}\right)^2 
\sum_{m,n}{n\,m\,\left[({\bf c}_n \cdot {\bf c}_m)({\bf d}_n \cdot {\bf
    d}_m)-({\bf c}_n \cdot {\bf d}_m)({\bf d}_n \cdot {\bf c}_m)\right]}\nonumber\\
&\le& \left(\frac {R^2 n_w^2}{\alpha'}\right)^2 
\sum_{m,n}{\frac {1}{4} n^2\,m^2 ({\bf c}_n^2 + {\bf d}_n^2) ({\bf
    c}_m^2 + {\bf d}_m^2)}\,.
\end{eqnarray}
Using the constraint equation (\ref{Virasoro}) with $m=0$, namely,
\beq
\sum_{n=1}^{\infty}{\frac {1}{2} n^2\, ({\bf c}_n^2 + {\bf d}_n^2)} = 1
\eeq
we arrive at 
\beq\label{j34bound}
|J_{34}|\le \frac {R^2 n_w^2}{\alpha'}\,, 
\eeq
or alternatively,
\beq\label{jbound}
|J_{34}| \le n_w J_{12}\,.
\eeq
Throughout the paper we take both $n_w$ and $J_{12}$ to be positively
oriented\footnote{The possibility of $n_w J_{12}<0$ would only make a
difference for heterotic strings. But the differences that appear are
largely irrelevant for our purposes.}.

It is clear that the vectors ${\bf c}_n$ and ${\bf d}_n$ could lie in
any spatial direction transverse to the 1-2 plane of the loop, and
therefore the construction generalizes immediately to any $D\geq 5$.
In this case we can have rotation in more independent planes.

Observe that the mass \reef{MRn} of the plasmid string is twice the energy
of a smooth static circular string wound $n_w$ times with the same
radius $R$. In other words, the amount of energy stored in the big loop
is the same as the energy carried by the small scale structure of the
string. This is, of course, not surprising, since it is this small-scale
structure that acts to balance the string tension at each point along
the string. In fact, we can easily derive the precise result from a
simple mechanical argument. If we consider a string with $n_p$ units of
momentum wound $n_w$ times on a circle of radius $R$, then its energy is
well-known to be
\beq\label{Enpnw}
E=\frac{n_p}{R}+\frac{R n_w}{\alpha'} \,. 
\eeq
In our construction the circle is contractible, so the
string that wraps it will not be in equilibrium for all values of $R$ but only for
those that extremize the energy, \ie those for which the `effective tension'
\beq\label{efftension}
\frac{dE}{dR}= -\frac{n_p}{R^2}+\frac{n_w}{\alpha'} \, 
\eeq
vanishes. This happens when
\beq\label{npnw}
\frac{n_p}{R} =  \frac{R n_w}{\alpha'}\,,
\eeq
which is the statement that the winding and momentum are `virialized'
and the total energy of the string is equally
divided into them\footnote{For strings
wound on a non-contractible circle, this corresponds to the self-T-dual
compactification radius.}. For our circular strings the
momentum along the string circle becomes rotation, so
\beq
n_p \to J_{12}\,,
\eeq
and we see that eqs.~\reef{Enpnw} and \reef{npnw} reproduce \reef{MRn}
and \reef{J12M}. Moreover, this argument shows that the circular string
is at a minimum of $E$ and so is {\it stable} to radial variations.

We conclude by noting that some of the states that we have constructed
are `special'. For instance, the plasmid \reef{plasmid} overlaps itself
whenever $N$ is a multiple of $n_w>1$. These states, however, do not
decay by breaking the string since on each wind the string is parallel
to itself. If $N$ is not a multiple of $n_w$, but $n_w>1$, the string
will not overlap but intersect itself, possibly several times, on each
turn, and this can lead to string breaking once interactions are switched
on. Still, if $n_w$ is very large the string will be almost parallel to
itself at the intersections. At any rate, we need not worry much about
these effects, since they will not be properties of generic string
states containing many small-scale wiggles.

\setcounter{equation}{0}
\section{Microscopic description of a circular string loop}
\label{sec:micro}

Our construction in the previous section was purely classical. It shows
that there is a large number of string configurations that, at large
scales, can be appropriately characterized as rotating loops of string.
However, in order to count the degeneracy of these circular strings we
must quantize the system. To this end, we turn to the full quantum
description of these configurations in string theory. We begin by
reviewing the procedure for defining the analogue of a coherent state
description of an extended classical closed string, following the
construction in \cite{BlancoPillado:2007hi} and generalizing it to allow
for $n_w>1$.

\subsection{Fixing the gauge and solving the constraints}

In light-cone gauge we set 
$X^+\equiv(X^0+X^{D-1})/\sqrt{2}= 2 \alpha' p_+\tau$. For the remaining 
coordinates we have
the standard decomposition for closed strings into left 
and right-movers, which on the solutions, takes the form
\begin{eqnarray}
X^i&=&X^i_L+X^i_R\,,\\
X^i_L&=&{1\over 2}x^i+\alpha^\prime p^i(\tau-\sigma)+
i\sqrt{\alpha^\prime\over 2}
\sum_{n\not = 0} {1\over n} \alpha_n^i{\rm e} ^{-2in(\tau-\sigma)}\,,\label{xl}\\
X^i_R&=&{1\over 2}x^i+\alpha^\prime p^i(\tau+\sigma)+i\sqrt{\alpha^\prime\over 2}
\sum_{n\not = 0} {1\over n} \tilde\alpha_n^i{\rm e}^{-2in(\tau+\sigma)}\,,
\end{eqnarray}
where $i=1, \dots, D-2$, and $x^i$ and $p^i$ are the center of mass
position and momentum of the string loop.

After reaching light-cone gauge there is still a residual symmetry:
rigid shifts of $\sigma$ generated by the operator $N_L-N_R$. Following
\cite{BlancoPillado:2007hi}, we fix this residual symmetry by adding a
suitable gauge fixing term to the action, 
\beq\label{Lgf}
{\cal L}_{gf}=\lambda \Phi\,,
\eeq
where
\beq
\Phi={1\over\pi}\int d\sigma\left({\rm e}^{-2in_w\sigma}\partial_-X^2
-n_wR\,{\rm e}^{-2in_w\tau}\right)\,.
\eeq
This allows us to solve the constraint $N_L-N_R=0$ and determine
$\alpha_{n_w}^2$. The parameters $R$ and $n_w$ will be the radius of the
circle and the winding number of the string loop respectively. Also note
that the gauge-fixing condition involves only the left-moving part of
$X^2$ since $\partial_- X_R\equiv 0$.

The gauge-fixing 
condition and the constraint $N_L-N_R=0$ can be solved (if $R\not =0$) for 
$\alpha_{n_w}^2$ and 
$\alpha_{-n_w}^2$, respectively, as 
\beq
\alpha_{n_w}^2&=&{n_wR\over\sqrt{2\alpha^\prime}}\,,\label{sol1}\\
\alpha_{-n_w}^2&=&-{\sqrt{2\alpha^\prime}\over n_wR}\left(\sum_{m\not =n_w} 
\alpha_{-m}^2
\alpha_m^2+\sum_{n\ge 1} \alpha_{-n}^j
\alpha_n^j-\t \alpha_{-n}^i\t\alpha_n^i\right)\,,\label{sol2}
\eeq
where $j=2,\dots,D-2$.

Upon quantization in this gauge, the mode decomposition of $X^2_L$ is
different from the usual one in (\ref{xl}). It reads
\beq\label{x1}
X^2_L&=&{1\over 2}x^2+\alpha^\prime p^2(\tau-\sigma)
- i\sqrt{\alpha^\prime\over 2}\, {1\over n_w}\alpha_{-n_w}^2{\rm e} ^{2in_w(\tau-\sigma)}
+ {i \over 2} R\, {\rm e} ^{-2in_w(\tau-\sigma)}\nonumber\\
&&+i\sqrt{\alpha^\prime\over 2}\sum_{n\not =n_w}{1\over 
n}\left(\alpha_n^2\,{\rm e}^{-2in(\tau-\sigma)}
-\alpha_{-n}^2\,{\rm e}^{ 2in(\tau-\sigma)}\right)\,,
\eeq 
where $\alpha_{-n_w}^2$ should be interpreted as the rhs of
(\ref{sol2}). Therefore, the Fock space of states does not include
$\alpha^2_{n_w}$ nor $\alpha^2_{-n_w}$ as operators acting on it. We
also note that since the constraint $N_L=N_R$ is solved for states in
this space, the
mass formula is
\beq
\alpha^\prime M^2=4N_R\,.
\eeq

\subsection{The state}

Let us now consider a state, in the gauge of the previous subsection, of the 
form
\beq\label{state0}
|\phi\rangle=|\phi_w\rangle_L\otimes|N_R\rangle_R\,,
\eeq
where the left-moving factor is a coherent state built on a left vacuum,
$|0\rangle_L$,
\beq\label{state}
|\phi_w\rangle_L=
{\rm e}^{-iR n_w \left(\alpha_{-n_w}^1+\alpha_{n_w}^1\right)/\sqrt{2\alpha^\prime}}|0\rangle_L\,,
\eeq
and the right-moving part is a state of level $N_R$. Level-matching requires that 
\beq\label{level}
N_R=\frac{n_w^2 R^2}{\alpha^\prime}\,.
\eeq  
Also, take vanishing center of mass parameters $x^i$ and $p^i$. Notice
that $\alpha^1_{n_w}|\phi_w\rangle_L=-\frac{in_wR}{\sqrt{2\alpha^\prime}}
|\phi_w\rangle_L$. Using (\ref{sol2}), this implies
$\langle\phi_w|\alpha^2_{-n_w}|\phi_w\rangle_L=
n_wR/\sqrt{2\alpha^\prime}$.

We can compute now the expectation value of the string coordinates in the
normalized state $|\phi\rangle$, and find the circular loop of radius
$R$ we seek,
\beq
\langle X^1\rangle&=& i\sqrt{\alpha^\prime\over 2}\langle\phi|
-{1\over n_w} \alpha_{-n_w}^1{\rm e}^{2in_w(\tau-\sigma)}+
{1\over n_w}\alpha_{n_w}^1{\rm e}^{-2in_w(\tau-\sigma)}|\phi\rangle\,,\nonumber\\
&=&R\,{\rm cos}\left( 2n_w(\tau-\sigma)\right)\,,\\
\langle X^2\rangle&=&i\sqrt{\alpha^\prime\over 2}\langle\phi|
-{1\over n_w}\alpha_{-n_w}^2{\rm e}^{2in_w(\tau-\sigma)}+ 
{R\over\sqrt{2\alpha^\prime}} {\rm e}^{-2in_w(\tau-\sigma)}|\phi\rangle\,,\nonumber\\
&=&R\, {\rm sin}\left( 2n_w(\tau-\sigma)\right)\,,
\label{x1x2}\eeq
in which the string winds the circle $n_w$ times before closing in on itself, so
that
\beq\label{nw}
\langle X^1\left(\sigma+\frac{\pi}{n_w}\right)\rangle=\langle
X^1(\sigma)\rangle\,,\qquad
\langle X^2\left(\sigma+\frac{\pi}{n_w}\right)\rangle=\langle
X^2(\sigma)\rangle\,.
\eeq

There is no contribution from the right-moving bosonic excitations to
the expectation value because we are considering that this sector is in
an eigenstate of $N_R$. Since $N_R=N_L$ is solved, $|N_R\rangle_R$ is
also an eigenstate of $N_L$. On the other hand, the coherent part
$|\phi_w\rangle_L$ is annihilated by both $N_R$ and $N_L$, and can be
regarded as a `background' on which we can put right-moving excitations.
This left-moving factor is the only one that produces the circular loop
of radius $R$ in the $X^1$-$X^2$ plane for a given mass.
Once $N_R$, and hence the mass, is fixed, adding a left-moving component
to this state would change the expectation value of the
rhs of \reef{sol2} and the shape of $X^2$ would change.

The mass and angular momentum on the state are
\beq\label{mandj}
\langle\phi |\alpha^\prime M^2|\phi\rangle &=& \langle\phi|4 N_R 
|\phi\rangle={4n_w^2R^2\over \alpha^\prime} \,,\\
\langle\phi |J_{12}|\phi\rangle &=&{n_wR^2\over \alpha^\prime} \,,\label{J} 
\eeq
where (\ref{J}) is computed on the left-moving part of the state using
\beq
J_{12}=-{i\over 2}\sum_{n\not =0}{1\over n}\alpha_{-n}^{[1}
\alpha_n^{2]}\,,
\eeq 
\ie we are considering only the left-moving part  
because on average the contribution to the ensemble of right-movers of a
given level has vanishing angular momentum. The result of dropping this
restriction is considered below in section~\ref{jr}. Therefore we obtain
the relations
\beq\label{JRM}
J_{12}=\frac{n_wR^2}{\alpha'}={\alpha^\prime\over 4n_w}M^2\,.
\eeq
These reproduce precisely the classical results \reef{MRn} and \reef{J12M}.

Our construction of a state that semiclassically resembles a plasmid
string has required a choice of gauge that is specific to the state we
are considering, which fixes the radius of the string in the direction
$X^2$. This has allowed us to have non-vanishing values for $\langle
X^1\rangle$ and $\langle X^2\rangle$ that coincide with the classical
values in \reef{plasmid}. However, one could still provide a different
construction of the circular string while leaving unfixed the residual
symmetry and imposing the constraint $N_L=N_R$ on physical states. This
can be done in such a way that we obtain the required values of $M$,
$J_{12}$, $n_w$, and with the root mean square position of the string
being peaked on a circle of a given radius. Consider the state
\beq\label{rcircle}
|\psi\rangle = |\psi\rangle_L \otimes |N_R\rangle_R =
\frac{1}{(J!~2^J)^{1/2}} (\alpha^1_{-n_w} + i
\alpha^2_{-n_w})^J |0\rangle_L \otimes|N_R\rangle_R\,.
\eeq
In this state we have
\begin{eqnarray}
J_{12}|\psi\rangle_L &=& J |\psi\rangle_L\,, \\
N_L |\psi\rangle_L &=& n_w J |\psi\rangle_L 
\end{eqnarray}
and $|N_R\rangle_R$ denotes an eigenstate of $N_R$.   
We must impose $N_R = n_w J$ in order to fulfill the 
level matching condition, $(N_R - N_L)|\psi \rangle = 0$. This implies  
\beq
\alpha' M^2 |\psi\rangle = 4 n_w J |\psi \rangle\,,
\eeq
so the classical relationship \reef{J12M} is satisfied
on the state, instead of only in expectation value like in
\reef{JRM}.
We know from \cite{BlancoPillado:2007hi} that any such 
state has $\langle X^i \rangle =0$, but one can measure 
the size of the string loop by the operator,
\beq
r^2  = (X^1)^2 + (X^2)^2 = \alpha' \sum_{n=1}^{\infty}
{\frac{1}{n^2} \left(\alpha^k_{-n} \alpha^k_{n} + \tilde \alpha^k_{-n}
  \tilde \alpha^k_{n}\right)}+ \alpha' \sum_{n=1}^{\infty}{\frac{1}{n}}~.
\eeq
where $k = 1,2$. The last term comes from reordering the operators to form the 
first sum, and gives a divergent contribution. This divergence was also noticed
in \cite{Karliner:1988hd}, where, given that it is independent of the
specific state, it was proposed that it be subtracted away. Doing the
same, we find that our state has
\beq
\langle r^2\rangle=\langle\psi| r^2 |\psi\rangle_L  = \frac{\alpha' J}{n_w}~.
\eeq
If we identify $R=\sqrt{\langle r^2 \rangle}$ then we reproduce the
classical relation \reef{J12nR}. Hence this state has the values of
$J_{12}$, $M^2$ and $R$ that we are seeking. We could also describe the
right-movers with a similar state, now in the 3-4 plane, thus yielding a
non-zero $J_{34}$ that reproduces the parameters of the classical state
\reef{plasmid}. 

The states \reef{state0} and \reef{rcircle} share the main macroscopic
parameters of the classical plasmid string, so both could be considered
to provide a quantum microscopic description of it. However, we feel
that the coherent state construction \reef{state0} captures more neatly
through eqs.~\reef{x1x2} the notion of a semiclassical rotating loop of
string.

\subsection{Degeneracies}

The coherent left-moving part in \reef{state0} is chosen to yield a
circular string with the required $n_w$ and with a fixed radius $R$ in
the direction $X^2$. Alternatively, we may say that we are fixing the
mass and this radius. The right-sector component is instead only
constrained to be at level $N_R$ given by \reef{level}. This allows a
large multiplicity for the string state. It is
straightforward to compute it in the limit of high level $N_R$ by studying
the appropriate generating function with standard
techniques \cite{Green:1987sp}\footnote{Clearly we will obtain the same asymptotic
degeneracy if instead of \reef{state0} we consider the state
\reef{rcircle}.}.

For bosonic strings ($D=26$, $i=1,\dots,  24$) we consider the partition function
for right-moving states with arbitrary occupation number $\{N_n^i\}$ for each 
mode. The level of each of these states is $N_R=\sum_i\sum_{n=1}n\,N_n^i$, 
so the generating function for this system is  
\beq\label{Zb}
Z_B={\rm tr} \,\e^{-\beta N_R}=\sum_{N_R=1}^\infty d_{N_R} \, z^{N_R}\,,
\eeq
where $z=\e^{-\beta}$ and $d_{N_R}$ is the degeneracy of states at
level ${N_R}$ and the trace is taken over the state space. A standard
saddle-point estimate yields
\beq
d_{N_R}\sim
\e ^{4\pi\sqrt{{N_R}}}\,, \qquad {N_R}\gg 1\,.
\eeq
In fact, the generic result for all closed string theories (bosonic,
IIA/B, heterotic) is
\beq
\log d_{N_R} \sim \sqrt{N_R}\,.
\eeq
The precise numerical factor, which varies among the theories, will not
be required for our purposes. Using \reef{level} and \reef{JRM} we find
that, to leading order at large $N_R$, the entropy of our
circular strings is
\beq\label{SJn}
S\sim \sqrt{n_wJ_{12}}\,.
\eeq

\subsection{Right-movers with angular momentum}
\label{jr}

We can also estimate the degeneracy of bosonic states when the 
right-movers have angular momentum in one direction, using the 
prescription and results of \cite {RS}.

We add a term containing the angular momentum and a Lagrange multiplier $\omega$ to the 
Hamiltonian of the right-movers,
\beq
H_J=\sum_{n,i} n N_n^i +\omega J_{34}\,, 
\eeq
where we chose the non-vanishing component of the angular momentum 
on the desired plane to be
\beq
J_{34}=-{i\over 2}\sum_{n\not =0}{1\over n}\t\alpha_{-n}^{[3}
\t\alpha_n^{4]}\,.
\eeq
The partition function of interest is now
\beq
Z_{J}={\rm tr\,e}^{-\beta H_J}\,.
\eeq
The saddle point approximation yields the degeneracy of states, which 
to leading order at large ${N_R}$ and $J_{34}$ is \cite{RS}
\beq
d_{{N_R},J_{34}}\sim 
\exp{\sqrt{4\pi(N_R-|J_{34}|)}}\,.
\eeq
For our states, the level is matched using \reef{level} as before. The
same result holds for all closed strings up to an overall numerical factor,
so the leading-order estimate for the entropy is
\beq\label{StwoJ}
S\sim \sqrt{{n_w^2R^2\over\alpha^\prime}-|J_{34}|}=\sqrt{n_wJ_{12}-|J_{34}|}.
\eeq
It is worth noting that at each level 
the number of states with nonvanishing $J_{34}$ is a subleading
fraction, proportional to $1/\sqrt{{N_R}}$, of those with $J_{34}=0$.

\setcounter{equation}{0}
\section{Supergravity solution for a rotating loop of fundamental string}
\label{sec:sugra}

We now want to find a solution that describes the supergravity fields
sourced by the rotating loop of fundamental string of the previous
sections. We first discuss the five-dimensional ring (so the
additional five space dimensions are assumed to be compactified, without
any non-trivial physics arising from the compactification), since in
this case exact explicit solutions are available. Afterwards we describe
the extension for arbitrary $D\geq 5$, where approximate solutions can
also be constructed.

\subsection{The 5D solution and its physical parameters}

The solutions we seek must describe a non-supersymmetric ring-like
object, with non-vanishing angular momentum, and possessing a dipole of
the Kalb-Ramond field $H_{(3)}$, but no other gauge charges. The wiggly
structure of the oscillations is assumed to be too small to be resolved
by the supergravity fields, and the degeneracy of the solutions is not
sufficient to give rise to a regular horizon, at least at the level of
the two-derivative supergravity action. So we expect a singularity on
the ring. 

Supergravity solutions with these properties were actually
presented in \cite{Emparan:2004wy}. In order to establish precisely
the identification we need to show that the relations \reef{JRM} and
\reef{SJn} are also satisfied by the supergravity solutions, at least in
the regime in which the string coupling constant is small. 
To this effect, we need to extract the physical parameters of the solution.
In the string frame, it takes the form\footnote{This
is obtained from the solutions in \cite{Emparan:2004wy} by setting $\nu=0$ (for
extremality), $N=1$ (for the dilaton coupling of fundamental
strings), 
$\lambda=\mu$ (for equilibrium), and finally changing to string frame.}
\beq\label{dipolering}
ds^2&=&-\frac{(1+\mu y)(1-\mu x)}{(1-\mu y)(1+\mu x)}
\left(dt+R\:\mu\sqrt{\frac{1+\mu}{1-\mu}}\: \frac{1+y}{1+\mu y}\: d\psi\right)^2\\[3mm]
&&+\frac{R^2}{(x-y)^2}\: (1-\mu^2 x^2)\left[
\frac{y^2-1}{1-\mu^2 y^2}d\psi^2+\frac{dy^2}{y^2-1}
+\frac{dx^2}{1-x^2}+\frac{1-x^2}{1-\mu^2 x^2}d\phi^2\right]\,,\nonumber
\eeq 
with Kalb-Ramond potential
\beq
B_{t\psi}=-R\:\mu\sqrt{\frac{1-\mu}{1+\mu}}\:\frac{1+y}{1-\mu y}\,, 
\eeq
and dilaton\beq
e^{2\phi}=\frac{1-\mu x}{1-\mu y}\,. 
\eeq
Readers unfamiliar with the set of coordinates employed here may find 
the explanation in \cite{Emparan:2006mm} helpful. Briefly,
$-R/y$, with $y\in (-\infty,-1]$, is a sort of radial coordinate in 
the direction away from the ring, and $x$ and $\phi$ parametrize 
two-spheres that link the ring once, with $x\sim \cos\theta$ and $x\in [-1,1]$. 
Regularity at the axes of $\phi$ and $\psi$ rotations requires the periodicities
$\Delta\phi=\Delta\psi=
2\pi \sqrt{1-\mu^2}$.
The solution has a (timelike) naked singularity at $y=-\infty$, where
the ring lies, and an
ergosurface around it, with topology $S^1\times S^2$, at $y=-1/\mu$. The
ring rotates along the $\psi$ direction.

There are two parameters, $R$ and $\mu$: $R$
corresponds to the ring radius and sets the scale for the solution. 
The dimensionless $\mu$ is related to the dipole of the $B_{(2)}$
field. This is proportional to the winding number of the string
obtained by integrating the flux of $H_{(3)}=dB_{(2)}$ across a 
2-sphere that links the string once (\ie any sphere at constant 
$y\in (-\infty,-1)$),
\beq
n_w&=&\frac{\alpha'}{8G}\int_{S^2}e^{-2\sqrt{\frac{2}{3}}\phi}\:\ast H_{(3)}
\nonumber\\
&=&\frac{\pi\alpha'}{2G}R\mu\,.
\eeq
We can use this equation to eliminate $\mu$ in favor of quantities with
direct physical meaning. In terms of these, the (Einstein frame) ADM
mass and angular 
momentum of the solution are
\beq\label{mass}
M=\frac{2 R}{\alpha'} n_w\left(1+\frac{G n_w}{\pi\alpha' R}\right)\,,
\eeq
\beq\label{angmom}
J_{12}=\frac{R^2}{\alpha'} n_w \left(1+\frac{2G n_w}{\pi\alpha' R}\right)^2\,. 
\eeq
We are denoting the plane where the ring rotates as
the 1-2 plane, to make contact with previous sections.

The presence of Newton's constant $G$ in these expressions is a sign of
the effect of self-gravitational interaction within the ring. These
effects are not included in the construction of the semiclassical string
state, so in order to make the comparison we must neglect them here too.
This requires that we consider `thin rings', such that
the `charge radius' $\mu R$ in directions transverse to the ring is
much smaller than the ring radius,
\beq\label{thinloop}
\frac{Gn_w}{\alpha'} \ll R\,.
\eeq
In effect, we
linearize the solution around flat space. Then, to zero-th order in 
$Gn_w/\alpha'R$ we have
\beq\label{linMJ}
M=\frac{2 R}{\alpha'} n_w+O(G)\,,\qquad J_{12}=\frac{R^2}{\alpha'} n_w+O(G)\,.
\eeq
To this order, these reproduce exactly \reef{JRM}. The supergravity
solution thus possesses the correct physical parameters to match the
semiclassical string loop. 

Let us analyze briefly
the effect of the $O(G)$ corrections. They appear when we account for
the fact that the ring attracts itself through interactions mediated by
gravitons as well as by $H_{(3)}$ and $\phi$ exchange. In the thin ring
regime \reef{thinloop} the interaction is at large distance so these
massless fields give rise to Newtonian and Coulombian forces. If we
expand \reef{mass} and \reef{angmom} beyond leading order, we find
\beq
M=2\sqrt{\frac{n_w J_{12}}{\alpha'}}-
\frac{G}{2\pi}\frac{M^2}{R^2}+ O(G^2)\,.
\eeq
The correction to the mass at first order in $G$ has indeed the form of
a Newtonian potential energy in five dimensions---in fact an attractive
one, as it should be since not only gravity but also the other fields
have a self-attractive effect on the ring.

Note that these corrections, which arise from closed string interactions
are {\it classical}. There will be other terms at the same order in the
string coupling with an interpretation as quantum corrections. They will
in fact give a decay rate for the state, to be discussed below.

\subsection{$D\geq 6$}
\label{sec:dgeq6}

In $D\geq 6$ we do not have explicit exact solutions for dipole rings, but
the methods of \cite{Emparan:2007wm} allow to construct
approximate solutions in the regime we are interested in, namely, to
first order in the parameter
\beq\label{thinring}
\frac{Gn_w}{\alpha' R^{D-4}} \ll 1\,.
\eeq
The method involves solving the supergravity equations in two regions,
first at distances $r\gg (Gn_w/\alpha')^{\frac{1}{D-4}}$, where the
linearized approximation around flat space is valid, and then near 
the ring core, $r\ll R$, where we perturb around the limit of a 
straight ($R\to\infty$) fundamental string with momentum. If 
\reef{thinring} holds, then there is an ample region where the 
two approximations are simultaneously valid and the respective 
solutions can be matched.

The construction is straightforward, if a little tedious, but we do not
need to develop it in full in order to derive the result we seek, namely
the relations \reef{JRM} ---which, observe, are independent of the
number of dimensions. We begin by considering the
solution near the ring to zero-th order in the parameter
\reef{thinring}. This is simply the solution for an extremal fundamental
string with momentum (FP-string), which in string frame is
\cite{Dabholkar:1995nc,Callan:1995hn}
\beq\label{fpstring}
ds^2=h^{-1}\left(-dt^2+dz^2+\frac{p}{r^{D-4}}(dt-dz)^2\right)+dr^2+r^2d\Omega^2_{D-3}
\eeq
with
\beq\label{hq}
h=1+\frac{q}{r^{D-4}}\,.
\eeq
The winding number is given by the charge $q$ as
\beq\label{nwq}
n_w
=\frac{(D-4)\Omega_{D-3}\alpha'}{8G}q\,,
\eeq
and if the direction $z$ is periodically identified $z\sim z+2\pi R$,
then the momentum parameter $p$ is quantized as
\beq\label{np}
\frac{n_p}{R}=\frac{(D-4)\Omega_{D-3}R}{8G}p\,.
\eeq

At large distances from the ring, $r\gg q^{1/(D-4)},\ p^{1/(D-4)}$, 
the gravitational field is the same as that created by a distributional
energy-momentum (measured in Einstein frame)\footnote{We normalize
$\int_{B^{D-2}}\delta^{(D-2)}(r)=\Omega_{D-3}$,
where $B^{D-2}$ is a ball that intersects the string once.}
\beq\label{tmunu}
T_{tt}&=&\frac{D-4}{16\pi G}\;(p+q)\;\delta^{(D-2)}(r)\,,\nonumber\\
T_{tz}&=&\frac{D-4}{16\pi G}\;p\;\delta^{(D-2)}(r)\,,\\
T_{zz}&=&\frac{D-4}{16\pi G}\;(p-q)\;\delta^{(D-2)}(r)\nonumber\,,
\eeq
plus a linear string source $\propto q$ for $H_{(3)}$ that we 
need not specify here. We now consider this same distributional source, but lying
along a ring of radius $R$ in $D$-dimensional flat spacetime,
parametrized by an angle $\psi=z/R \in [0,2\pi)$. Then the momentum $p$
along $z$ becomes proportional to an angular momentum $J_{12}$ along
$\psi$, and the charge $q$ becomes the dipole of the ring. The
linearized supergravity equations can be solved for this source, 
providing a solution valid at distances $\gg
(Gn_w/\alpha')^{\frac{1}{D-4}}$. It is not difficult to
do this, but in order to extract the physical parameters of the ring we
need only notice that at any finite $R$ the ring will be in mechanical
equilibrium only when its tendency to collapse under its tension is
balanced by the centrifugal repulsion. The condition for this to happen
is that
\beq\label{equilt}
\frac{T_{zz}}{R}=0\,.
\eeq
This was argued in \cite{Emparan:2007wm} from several points of view: it
follows from the classical equation of motion for a probe brane, derived
in \cite{Carter:2000wv} as a consequence of conservation of the
stress-energy tensor. Perhaps more appropriately for our present
purposes, ref.~\cite{Emparan:2007wm} showed that it
follows from the requirement of absence of singularities on the plane of
the self-gravitating thin ring, away from the ring location. Both
arguments lead to \reef{equilt} in the present
case. Imposed on \reef{tmunu} it implies that the ring will
be in mechanical equilibrium only when
\beq\label{pisq}
p=q\,.
\eeq
Using \reef{nwq} and \reef{np}, which remain valid to this order of
approximation, this gives the mass and angular momentum of the ring as
\beq
M&=&2\pi R \int_{B^{D-2}}T_{tt}=\frac{(D-4)\Omega_{D-3}}{4G} q R=\frac{2R}{\alpha'}n_w\,,\\
J_{12}&=&2\pi R^2 \int_{B^{D-2}}T_{tz}=
n_p=\frac{R^2}{\alpha'}n_w\,.
\eeq
So we find that the expressions \reef{JRM} are again exactly reproduced.
Hence thin extremal dipole rings provide the correct supergravity
description of semiclassical circular loops of string in any $D\geq 5$.
It may be worth noting that the integrated expressions for $T_{tt}$ and
$T_{zz}$ correspond to the energy and effective tension introduced in
\reef{Enpnw} and \reef{efftension}, and so \reef{pisq} is clearly the
same as \reef{npnw}.

Arguments of the sort discussed in the previous subsection indicate that
the first $O(G)$ corrections away from the thin ring limit will reduce the
mass by an amount $\propto GM^2/R^{D-3}$.

Finally, note that the supergravity solution for a fundamental string
with only fermionic excitations should take the same form as
\reef{fpstring}, so our construction also applies to the
circular strings of \cite{BlancoPillado:2005px}.

\subsection{Entropy}

The geometry \reef{dipolering} does not have a horizon, but instead a
naked singularity, so it would seem that there is no entropy associated
to it. How can we then identify it with the perturbative string state,
which has a degeneracy \reef{SJn}? The resolution is that the entropy
\reef{SJn} is too small to show up as a macroscopic entropy in the
leading low-energy effective supergravity description \cite{Sen:1995in}. The
situation is in fact closely analogous to the `small black holes' that
correspond to elementary string states wrapped on a circle. In recent
years it has been argued that the inclusion of higher-derivative
corrections to the supergravity action removes the singularity at the
core and replaces it with a `stretched' horizon. The Wald entropy of this horizon
precisely reproduces the microscopic entropy of the fundamental string
state \cite{Dabholkar:2004yr}.

It turns out that we can immediately apply to our rotating rings the
same arguments and results that \cite{Sen:1995in,Sen:2005kj} developed
for the straight fundamental string. The reason is that in the thin ring
limit \reef{thinloop} the near-horizon geometry
is exactly the same as that of the straight string, namely
eq.~\reef{fpstring}, now with $h=q/r^{D-4}$. Deviations from this
near-horizon geometry come from self-interaction of the loop, which we
are neglecting. Ref.~\cite{Sen:2005kj} developed a scaling argument,
based only on this near-horizon geometry, to the effect that the entropy
of the stretched horizon is
\beq
S\sim \sqrt{n_w n_p},
\eeq
where $n_p$ are the units of momentum
along the string. This is valid in any dimension $D\geq 5$. When we bend the string to form a
circle, we have $n_p\to J_{12}$, so
\beq\label{Sscale}
S\sim \sqrt{n_w J_{12}},
\eeq
which reproduces the parametric dependence \reef{SJn} of the microscopic
string state. Fixing the precise factor requires control over
higher-derivative corrections, which is currently unavailable for
generic $D$. However, the scaling \reef{Sscale} is a robust result.

\subsection{Second angular momentum: Tubular loops}

We have seen that it is possible for the right-movers on the string to
contribute a second angular momentum $J_{34}$ in a plane orthogonal to
the plane of the loop. This angular momentum is bounded above by
\reef{jbound}. It is now natural to ask what is the supergravity counterpart
of these solutions.

There do exist some exact solutions for black rings with two independent
angular momenta \cite{Elvang:2004rt,Elvang:2004xi,Pomeransky:2006bd},
and they all satisfy a bound $|J_{34}|<J_{12}$. None of these, however, is
a black ring with a dipole of $H_{(3)}$ and no other gauge charge. We
will argue that, in fact, the supergravity solution we seek here is not
a ring but instead a torus --- \ie not a loop of string but a loop of
tube, or {\it tubular loop}.

Begin by considering the solutions in the limit that the radius of the
loop $R\to \infty$. In this limit the solutions are BPS, so they should
correspond to supersymmetric rotating FP strings. Such solutions are
well-known to be helical strings that in the supergravity description are
smeared along the direction of the axis and thus describe a tube
$S^1\times {\bf R}$---this is the topology of the locus where the
spatial section of the supertube worldvolume lies. The exact
solutions have been presented in \cite{Lunin:2001fv}. We
now want to bend the tube axis into a circle of radius $R$, so the result
will be a toroidal $S^1\times S^1$ tube, see
fig.~\ref{fig:loopytube}\footnote{The helical strings of
\cite{Lunin:2001fv} are U-dual to the supergravity supertubes
of \cite{Emparan:2001ux}. However, T-duality along the loop is not a valid
symmetry among tubular loops
(it has fixed points at the axis). Since the D0-F1/D2
supertube of \cite{Mateos:2001qs,Emparan:2001ux} does not carry momentum along the tube
direction, it will not, if bent to
form a loop, possess the centrifugal motion to balance its tension.}. 
\begin{figure}
\centerline{\includegraphics[width=8cm]{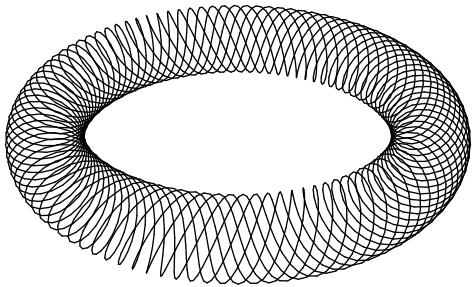}}
\caption{\small Tubular loop: when the radius of the helix oscillations is
macroscopically large, instead of a loop of string we obtain a toroidal
tube (for large $n_w$ the string is very uniformly spread on the tube;
in the picture $n_w=7$). It rotates in the two circles of the torus,
with $|J_{34}|\leq n_w J_{12}$.}
\label{fig:loopytube}
\end{figure}

When $R$ is large, an
approximate supergravity solution (which breaks all supersymmetries) can
be constructed, in any $D\geq 5$, using the methods of
\cite{Emparan:2007wm}. Just like in sec.~\ref{sec:dgeq6}, we do not
actually need to perform this construction explicitly in order to
extract the physical parameters in the limit where the system is weakly
interacting with itself. It has been shown that a supergravity supertube
rotating in the 3-4 plane satisfies the bound \cite{Lunin:2001fv,Emparan:2001ux}
\beq 
|J_{34}|\leq n_w n_p\,. 
\eeq 
When we bend the tube on the plane 1-2 to form the torus we identify
$n_p=J_{12}$. Moreover, it is easily seen that mechanical equilibrium of
the tube, \ie the condition that $T_{zz}=0$, requires again that \reef{npnw}
holds. Thus we recover \reef{j34bound} and \reef{jbound}.

The string state has a microscopic degeneracy given by \reef{StwoJ},
whose origin is the possibility of a wiggly structure that `thickens' the
tube. In turn, the supergravity solution for the supertube develops a
stretched horizon with an entropy that, in any
$D\geq 5$, is
\beq\label{Stube}
S \sim \sqrt{n_w n_p -|J_{34}|}\,.
\eeq
This has been derived in
\cite{Dabholkar:2006za} using the scaling arguments of
\cite{Sen:2005kj}. In the thin tube limit we can apply this formula to
our tubular loop, with
the substitution $n_p\to J_{12}$. Then the entropy \reef{Stube}
\beq
S \sim \sqrt{n_w J_{12}-|J_{34}|}\,
\eeq
reproduces  \reef{StwoJ}.
Observe that the topology of the stretched horizon is in this case
$S^1\times S^1 \times S^{D-4}$. Angular momenta on other planes would
blow up new circles.

\subsection{String decay by superradiant emission}

The quantum state of the fundamental string \reef{state0}, \reef{state},
contains excitations in both left and right sectors, so it is not a BPS
state and should have a finite lifetime when interactions are switched
on. 

On the other hand, the supergravity solution for the ring
\reef{dipolering} is also expected to decay. The decay is unlikely to
happen at the classical level, since this ring is not affected by any of
the black ring instabilities discussed in \cite{Elvang:2006dd}. But
quantum decay is certainly expected: not by Hawking emission, since the
solution is extremal and has zero temperature\footnote{This is properly
defined after the near-horizon geometry is regularized with
higher-derivative corrections to an AdS type of horizon.}, so it does
not emit thermal radiation. But it has an ergoregion. Extremal black
holes with an ergoregion surrounding a horizon with angular
velocity $\Omega_H$, do emit spontaneously modes $\Psi \sim e^{-i(\omega t
-m \phi)}$ that satisfy a superradiant bound on their frequency $\omega
< m \Omega_H$ \cite{Page:1976ki}. This emission carries away some of the
angular momentum of the black hole. The calculation of this process in
the background of the ring \reef{dipolering} is technically complicated,
since the variables $x$ and $y$ cannot be separated in the wave
equation\footnote{In principle the horizon needs to be regularized by
higher-derivative corrections, but presumably a simple absorptive
boundary condition at the stretched horizon is enough to derive
superradiance.}. Nevertheless, on general grounds we expect that this
ring, as well as its $D>5$ counterparts, will decay by such superradiant
emission. 

It is possible to identify the counterpart of this radiation in the
microscopic model. The main ideas are contained in the microscopic
picture of black hole superradiance developed in
\cite{Emparan:2007en,DEM} using the microscopic dual of an extremal
rotating black hole with an ergosphere. This system possesses
essentially the same features as our configurations. In both cases we
have a state of a 1+1 CFT where the left-moving sector is filled with
coherently polarized excitations so it accounts for the angular rotation
of the horizon and is at zero temperature. The right-moving sector,
instead, is in a thermal ensemble and accounts for the entropy, and
possibly for angular momentum in an orthogonal plane. The state can
decay by an interaction between the two sectors, which results in the
emission of a massless closed string. The left-moving excitation
provides angular momentum, so the emitted quantum will carry away some
spin from the system. In our case, the emission will reduce the value of
$R$, so the radius of the loop, and the angular momentum $J_{12}$, will
decrease. It is natural to expect that this decay corresponds to the
superradiant emission of the black ring. 

For the extremal rotating black hole in \cite{DEM}, it is possible to
show that the microscopic model implies that the emitted quanta do
indeed satisfy the bound $\omega < m \Omega_H$. It should be interesting
to also match, at least parametrically, the superradiant frequency bound
for our circular strings using their micro and macro descriptions. Note
that the decay of a large semiclassical state is expected to be slow, so
these circular strings should be long-lived. In addition, as a result of
the decay, the left sector of the string will gradually lose coherence.
In the supergravity side, the temperature will be raised from zero and
the black ring will become non-extremal.

Finally, notice also that the classical plasmid string of
section~\ref{sec:classical}, or a generic state with arbitrary wiggly
profile, has a varying mass-quadrupole moment so when it is coupled to
gravity it is expected to radiate gravitational waves\footnote{The
decay of the fundamental string through emission of classical
gravitational waves is considered in \cite{Iengo:2006gm}.}. However, this
radiation is strongly suppressed for large $N$, \ie when the wiggle
amplitude is small and rotational invariance along the loop is
approximately recovered. We expect that the superradiant emission in the
supergravity solution (which is rotationally invariant) can be set in
correspondence with this suppressed classical radiation.

\setcounter{equation}{0}
\section{Discussion}
\label{sec:discuss}

The main thread of the paper has been the idea that black rings can be
regarded as circular strings. This is a different perspective than
viewing them as supertubes dimensionally reduced along the tube
direction, as first proposed in \cite{Elvang:2003mj}. In the context of
states of the fundamental string, the latter view is taken in
\cite{Dabholkar:2005qs,Dabholkar:2006za}, where certain BPS states of
the string are related to a class of small black rings. These rings are
the helical strings, smeared along the helix axis, of
\cite{Lunin:2001fv}, which are U-dual to supergravity supertubes
\cite{Emparan:2001ux}. Dimensional reduction along the axis direction
yields a two-charge ring that saturates a BPS bound. The configurations
we have discussed are different, and in some respects simpler, than
these. They are not supersymmetric since they only have dipoles, not
conserved charges, hence they are more similar to neutral black holes.
Even when we have discussed supertubes, we have broken their
supersymmetry to form a tubular loop, not a ring.

The limitation to thin rings in our correspondence between the different
descriptions of the circular string may look like an important
deficiency compared to what can be achieved for supersymmetric black
rings. However, we believe that the understanding we have obtained of the
properties of black rings, when viewed as circular strings, is
significant. In particular, we have in mind a main drawback of the
otherwise successful identification of the microstates of supersymmetric
black rings \cite{Elvang:2004rt} when regarded as circular versions of
the MSW string \cite{Cyrier:2004hj,Bena:2004tk}. This model does not
appear to account for the fact that the ring wraps a contractible cycle,
and that therefore its radius is fixed in terms of parameters such as
the angular momentum and mass. Our analysis of dipole rings does
precisely this, and may provide hints about what is missing in the
supersymmetric ring description of \cite{Cyrier:2004hj,Bena:2004tk}.
Besides, our picture may form the basis for a better understanding of
the microphysics of neutral black rings, following the suggestions in
\cite{Emparan:2004wy}.

Our discussion of the regularization of the stretched horizon has been
based on the expectation that higher-derivative corrections will reveal
a small AdS-type horizon \cite{Dabholkar:2004yr,Sen:2005kj}. However,
there is another perspective on this issue, suggested in
\cite{Lunin:2002qf} as a prototype for the `fuzzball' proposal: the FP
string solution \reef{fpstring} is only an effective, coarse-grained
geometry for more fundamental solutions that are in one-to-one
correspondence to individual horizonless string states. Such
supergravity microstates are known explicitly for the straight string
\reef{fpstring}, and generically describe a wiggly pattern of
oscillating waves along the string. In this spirit,
we should find geometries for each of the classical string
configurations in sec.~\ref{sec:classical}. Solving exactly the supergravity
equations to find these configurations is too difficult, but one
might try the method of matched asymptotic expansions of
\cite{Emparan:2007wm} to build an approximate solution in the thin ring
regime \reef{thinring}. Presumably this is still technically
challenging, but we can anticipate an important feature: since the
wiggles break the exact rotational invariance of the ring, they will
give rise to a varying quadrupole and the configuration will radiate
gravitational waves---just like we argued at the end of
sec.~\ref{sec:sugra} for the classical plasmid string. These waves will
carry not only mass but also angular momentum away from the ring.
However, the emission will be suppressed for small wiggles. The coarse-grained
geometry \reef{dipolering} is rotationally invariant and does not emit
classical radiation, but it decays through superradiant emission, so,
again, it is natural to conjecture that both decays are dual
descriptions of the same phenomenon. In a very recent paper
\cite{Chowdhury:2007jx}, a closely similar emission from a set
of microscopic states is also put in correspondence with macroscopic
superradiant decay.

Finally, our arguments and constructions have all referred to $D\geq 5$,
since the four-dimensional case presents peculiarities of its own.
First, the semiclassical construction of section~\ref{sec:classical} in
terms of purely bosonic wiggly excitations would require that the string
oscillates as well in at least one of the 1-2 directions, and this
typically leads to cusps. Still, this problem is avoided if the
excitations are purely fermionic, see \cite{BlancoPillado:2005px}. The
approximate supergravity construction described in sec.~\ref{sec:sugra}
seems to extend as well to $D=4$, in spite of the logarithmic behavior
of the solution near the string core. However, in $D=4$ the scaling
argument of \cite{Sen:2005kj} fails to produce a regular horizon of size
parametrically larger than the string length \cite{Iizuka:2007sk},
essentially because $q$ and $p$ are dimensionless and there is no scale
for the horizon radius other than the string length. The conclusion
seems to be that four-dimensional loops of string are possible, but they
give rise to supergravity rings with string-scale cores and not (small)
black rings. It may be interesting to further investigate this system.

\section*{Acknowledgements} 
We would like to thank Jaume Garriga, Ken Olum, Jorge Russo and Alexander 
Vilenkin
for many useful discussions. AI would like to thank the CERN-TH Division
and the Department of Fundamental Physics of Universitat de Barcelona
for their hospitality while this work was in progress. J.J.B-P. was
supported by the National Science Foundation Grant PHY-06533561. RE was
supported in part by DURSI 2005 SGR 00082, CICYT FPA 2004-04582-C02-02,
and the European Community FP6 program MRTN-CT-2004-005104. AI was
supported by DOE Grant DE-FG03-91ER40674.

\end{document}